\documentstyle[12pt]{article}
\begin{document}

\title{{\bf Quasi degenerate Neutrino Masses with Universal Strength Yukawa
Couplings}}
\author{\vspace*{1em} \\
%EndAName
{\bf G. C. Branco}\thanks{%
E-mail address : d2003@beta.ist.utl.pt}, ~{\bf M. N. Rebelo}\thanks{%
E-mail address : rebelo@beta.ist.utl.pt} ~and ~{\bf J. I. Silva-Marcos}%
\thanks{%
E-mail address : juca@cfif1.ist.utl.pt}\\
CFIF/IST and Departamento de F\'{\i}sica\\
Instituto Superior T\'ecnico\\
Av. Rovisco Pais,\\
P-1096 Lisboa Codex, Portugal\\
\\
}
\date{}
\maketitle

\begin{abstract}
A simple ansatz is proposed for neutrino and charged lepton mass matrices,
within the framework of universal strength for Yukawa couplings. In this
framework all Yukawa couplings have equal moduli and the flavour dependence
is only in their phases. We take into account the solar neutrino deficit and
the atmospheric neutrino anomaly, assuming three neutrino families only. The
ansatz leads in a natural way to small mixing involving neutrinos of quasi
degenerate masses, as required to explain the solar neutrino deficit in the
non-adiabatic MSW solution, while having the large mixing necessary to
explain the atmospheric neutrino anomaly.
\end{abstract}

\setlength{\baselineskip}{14pt} 
\renewcommand{\thesubsection}{\arabic{subsection}} 
\begin{picture}(0,0)
       \put(335,475){FISIST/2-98/CFIF}
       \put(335,460){}
\end{picture}
\vspace{-24pt} \thispagestyle{empty} 
%%%%%%%%%%%%%%% abstract %%%%%%%%%%%%%%%%%%%%%%%%%

%%%%%%%%%%%%%%%%% Intro %%%%%%%%%%%%%%%%%%%%%%%%

\section{\bf Introduction}

%%%%%%%%%%%%%%%%%%%% 1 %%%%%%%%%%%%%%%%%%%%%%%%%%%%%%%%%

Presently the Standard Model (SM) enjoys quite a remarkable success when
confronted with experiment. However, there is recent experimental evidence
pointing towards physics beyond the SM in the leptonic sector, to wit, the
solar neutrino deficit, the atmospheric neutrino problem, and the results of
the LSND collaboration suggesting that neutrino oscillations might have been
observed in an accelerator experiment.

The solar neutrino data obtained by several different experiments $\cite{re1}
$, indicate a deficit in the number of observed neutrinos by comparison to
the Standard Solar Model (SSM) predictions for the solar neutrino fluxes $%
\cite{re2}$. The solar neutrino deficit is explained in terms of
oscillations of the electron neutrino into some other neutrino species. In
the framework of the MSW mechanism $\cite{re3}$ there are two sets of
solutions, the adiabatic branch (AMSW) requiring a large mixing $(\sin
^22\theta \simeq 0.65\sim 0.85)\cite{re4}$ and the non adiabatic branch
(NAMSW) requiring a small mixing $(\sin ^22\theta \simeq (0.1\sim 2)\times
10^{-2})\cite{re4}$ with $\Delta m^2\simeq (0.3\sim 1.2)\times 10^{-5}eV^2$.
The small mixing solution seems to be favoured by the present data. In the
framework of vacuum oscillation only three separate regions within a narrow
range of parameters ($\Delta m^2=5-8\times 10^{-11}eV^2$ and $\sin ^22\theta
=0.65-1$)$\ $are allowed $\cite{re4}$.

Several experiments have measured the ratio of the number of muon neutrinos
by the number of electron neutrinos produced in the atmosphere through the
decay of pions and kaons with subsequent decay of secondary muons $\cite
{re4a}$. The combined results lead to, 
\begin{equation}
\label{eq0}R\equiv \frac{(n_{\nu _\mu }/n_{\nu _e})_{{\rm {Data}}}}{(n_{\nu
_\mu }/n_{\nu _e})_{{\rm {SM}}}}\simeq 0.6 
\end{equation}
where $(n_{\nu _\mu }/n_{\nu _e})_{{\rm {Data}}}$ is the measured ratio of
muon-neutrino to electron-neutrino events, while $(n_{\nu _\mu }/n_{\nu
_e})_{{\rm {SM}}}$ is the expected ratio assuming no oscillations. This
anomaly can be caused by oscillations of the atmospheric muon neutrinos into
another type of neutrino with large mixing angle $(\sin ^22\theta \simeq
0.6\sim 1.0,\Delta m^2\simeq (0.3\sim 3)\times 10^{-2}eV^2)$.

The LSND group has reported evidence for $\overline{\nu _\mu }-\overline{\nu
_e}$ and $\nu _\mu -\nu _e$ oscillations $\cite{re5}$. However no other
accelerator experiment confirmed their result thus strongly reducing the
allowed parameter space. The resulting experimental constraints on neutrino
masses and mixings are such that no combined solution to the solar,
atmospheric and LSND results can be found in the framework of three
neutrinos only, since the required neutrino mass differences do not satisfy
the relation $\Delta m_{21}^2+\Delta m_{32}^2=\Delta m_{31}^2$. We have
chosen to consider simply three neutrino families without additional sterile
neutrinos and not to take into consideration the LSND data.

Astrophysical considerations, in particular the possibility that neutrinos
constitute the hot dark matter, favour neutrino masses of the order of a few 
$eV$ $\cite{re5a}$, which combined with the above constraints leads to a set
of highly degenerate neutrinos.

In this paper we propose a simple ansatz within the framework of universal
strength for Yukawa couplings (USY) $\cite{re6}$ which leads to
quasi-degeneracy of neutrino masses and provides a solution to the solar and
atmospheric neutrino problems. In USY all Yukawa couplings have equal moduli
so that the flavour dependence is only contained in their phases. For the
quark sector it has been shown $\cite{re7},\cite{re8}$ that the USY
hypothesis leads to a highly predictive and successful ansatz for quark
masses and mixings. Various mixing schemes for the leptonic sector have been
suggested $\cite{rem}$. The possibility of quasi-degenerate neutrino masses
has been recently considered in the literature in the context of some
specific symmetry or ansatz. An \underline {important feature} of our ansatz
lies on the fact that it can accommodate in a natural way small mixing
involving a set of highly degenerate neutrinos in the MSW solution to the
solar neutrino problem while also having the large mixing necessary to
explain the atmospheric neutrino data.

%%%%%%%%%%%%%%%%%%%%%%%%%%%%%%%%%%%%%%%%

\section{\bf Degeneracy in USY}

The USY hypothesis leads to neutrino mass matrices of the form:

\begin{equation}
\label{eq1}
\begin{array}{c}
M_\nu =c_{\nu}\ \left[ \ e^{i\theta _{ij}\ }\right] 
\end{array}
\end{equation}
where $c_{\nu}$ is an overall constant. Let us derive the conditions which
should be satisfied so that the matrices of Eq.(\ref{eq1}) lead to at least
two degenerate neutrinos. It is useful to introduce the dimensionless
Hermitian matrix $H_\nu\equiv M_\nu M_\nu ^{\dagger }\ /\ 3{c_\nu}^2$ which
can be written as:

\begin{equation}
\label{eq2}H_\nu =\left[ 
\begin{array}{lll}
1 & r_1e^{i\varphi _1} & r_2e^{i\varphi _2} \\ 
r_1e^{-i\varphi _1} & 1 & r_3e^{i\varphi _3} \\ 
r_2e^{-i\varphi _2} & r_3e^{-i\varphi _3} & 1 
\end{array}
\right] 
\end{equation}
where the off diagonal elements $r_ie^{i\varphi _i}$ are the sum of products
of pure phase elements of $M_\nu $:

\begin{equation}
\label{eq3}
\begin{array}{c}
(H_\nu )_{12}=r_1e^{i\varphi _1}={\frac 13\ }\left[ e^{i(\theta _{11}-\theta
_{21})}+e^{i(\theta _{12}-\theta _{22})}+e^{i(\theta _{13}-\theta
_{23})}\right] 
\end{array}
\end{equation}
with analogous expressions for $(H_\nu )_{13}$ and $(H_\nu )_{23}$. It can
be readily verified that if at least two of the eigenvalues of $H_\nu $ are
equal, then the following relation holds:

\begin{equation}
\label{eq4}
\begin{array}{c}
\left[ 1-{\frac \chi 3}\right] ^3=\left[ 1-{\frac{\chi -\delta }2}\right] ^2 
\end{array}
\end{equation}
where:

\begin{equation}
\label{eq5}
\begin{array}{ll}
\delta & \equiv \det (H_\nu )=\lambda _1\lambda _2\lambda _3 \\ 
\chi & \equiv \chi (H_\nu )=\lambda _1\lambda _2+\lambda _1\lambda
_3+\lambda _2\lambda _3 
\end{array}
\end{equation}
and the $\lambda _i=3m_i^2/(m_1^2+m_2^2+m_3^2)$ denote the eigenvalues of $%
H_\nu $. In the derivation of Eq.(\ref{eq4}) we took into account that $%
tr(H_\nu )\equiv \lambda _1+\lambda _2+\lambda _3=3$. The invariants $\chi $
and $\delta $ can be expressed in terms of $r_i$ and $\varphi _i$, with all
generality, as:

\begin{equation}
\label{eq6}
\begin{array}{ll}
\chi & =3-r_1^2-r_2^2-r_3^2 \\ 
\delta & =1+2\ r_1r_2r_3\cos (\varphi _1+\varphi _3-\varphi
_2)-r_1^2-r_2^2-r_3^2 
\end{array}
\end{equation}

The following ``spherical'' parametrization for the $r_i$ is useful:

\begin{equation}
\label{eq7}
\begin{array}{lll}
r_1= & [3(1-\chi /3)]^{1/2} & \sin \theta \ \cos \phi \\ 
r_2= & [3(1-\chi /3)]^{1/2} & \sin \theta \ \sin \phi \\ 
r_3= & [3(1-\chi /3)]^{1/2} & \cos \theta 
\end{array}
\end{equation}
with $0\leq \theta ,\phi \leq {\frac \pi 2}$. From Eqs.(\ref{eq6}) and (\ref
{eq7}), one obtains:

\begin{equation}
\label{eq8}
\begin{array}{c}
\sin ^2\theta \cos \theta \cdot \sin (2\phi )\cdot \cos (\varphi _1+\varphi
_3-\varphi _2)={\frac 2{3\sqrt{3}}\ }\left[ 1-{\frac{\chi -\delta }2}\right]
/\left[ 1-{\frac \chi 3}\right] ^{\frac 32} 
\end{array}
\end{equation}

In the case of at least two degenerate neutrinos Eq.(\ref{eq8}) can be
combined with Eq.(\ref{eq4}) leading to the condition:

\begin{equation}
\label{eq9}
\begin{array}{c}
\sin ^2\theta \cos \theta \cdot \sin (2\phi )\cdot \cos (\varphi _1+\varphi
_3-\varphi _2)={\frac 2{3\sqrt{3}}} 
\end{array}
\end{equation}
which can only be satisfied for $\cos \theta =1/\sqrt{3}$, $\sin (2\phi )=1$
and $\cos (\varphi _1+\varphi _3-\varphi _2)=1$. Therefore, the necessary
and sufficient conditions for $H_\nu $ to have two degenerate eigenvalues
are:

\begin{equation}
\label{eq10}
\begin{array}{l}
\varphi _1+\varphi _3-\varphi _2=0~( 
{\rm {mod}.2\pi )} \\  \\ 
r_1=r_2=r_3 
\end{array}
\end{equation}

Within the USY framework, it can be shown that for matrices $M_\nu $ having
at least two degenerate eigenvalues, there is a weak-basis where $M_\nu $
has one of the following forms (modulo trivial permutations) 
\begin{equation}
\label{eq11}
\begin{array}{ll}
M_\nu ^{{\rm I}}=c_\nu \ K\cdot \left[ 
\begin{array}{lll}
e^{i\alpha } & 1 & 1 \\ 
1 & e^{i\alpha } & 1 \\ 
1 & 1 & e^{i\alpha } 
\end{array}
\right] ; & M_\nu ^{{\rm I\!I}}=c_\nu \ K\cdot \left[ 
\begin{array}{lll}
e^{i\alpha } & 1 & 1 \\ 
1 & e^{i\alpha } & 1 \\ 
1 & 1 & e^{-2i\alpha } 
\end{array}
\right] \\  
&  \\ 
M_\nu ^{{\rm I\!I\!I}}=c_\nu \ K\cdot \left[ 
\begin{array}{lll}
e^{i\alpha } & 1 & 1 \\ 
1 & -1 & 1 \\ 
1 & 1 & -1 
\end{array}
\right] &  
\end{array}
\end{equation}
where $K\equiv $diag$(e^{i\varphi _1},e^{i\varphi _2},e^{i\varphi _3})$ and $%
c_\nu $ is a real constant.

The first two cases of Eq.(\ref{eq11}) are of special interest since for $%
\alpha =2\pi /3$ they lead to three degenerate neutrino masses, while for $%
\alpha \not =2\pi /3$ they lead to only two degenerate mass eigenvalues.

\section{\bf A Special Ansatz}

In order to have an ansatz with predictions for the leptonic mixing matrix,
one has to specify the structure of the charged lepton mass matrix $M_{\ell}$
together with the structure of $M_{\nu}$.

Our guiding principle is the assumption that all leptonic Yukawa couplings
obey the USY hypothesis. Furthermore, we choose, within USY, the same
structure of phases for the charged lepton and neutrino mass matrices.
Guided by the above ideas we propose the following specific ansatz:

\begin{equation}
\label{eq12}
\begin{array}{ccc}
M_\ell =c_\ell \left[ 
\begin{array}{lll}
e^{-ia} & 1 & 1 \\ 
1 & e^{ia} & 1 \\ 
1 & 1 & e^{ib} 
\end{array}
\right] & ; & M_\nu =c_\nu \left[ 
\begin{array}{lll}
e^{i\alpha } & 1 & 1 \\ 
1 & e^{i\alpha } & 1 \\ 
1 & 1 & e^{i\beta } 
\end{array}
\right] 
\end{array}
~~ 
\end{equation}
with $c_\ell $, $c_\nu $ real constants.

The leptonic mixing matrix $V$ appearing in the charged weak current is then
given by

\begin{equation}
\label{eq13}
\begin{array}{c}
V=U_\nu ^{\dag }\cdot U_\ell 
\end{array}
\end{equation}
where the matrix $U_\ell $ diagonalizes $M_\ell \ M_\ell ^{\dagger }$ and
the matrix $U_\nu $ diagonalizes $M_\nu \ M_\nu ^{\dag }$ in the case of
Dirac neutrinos \footnote{%
If both Dirac mass terms ($M_D$) and Majorana mass terms for the righthanded
neutrinos $(M_R)$ are present, with the light Majorana neutrinos acquiring
mass through the seesaw mechanism $\cite{see}$, we identify $M_\nu $ of
Eq.(12) with the effective mass matrix for the light neutrinos given by $%
M_\nu =M_DM_R^{-1}M_D^T$ and, in this case, $U_\nu $ is very approximately
the matrix verifying $M_\nu =U_\nu \cdot $diag$(m_{{\nu }_1},m_{{\nu }_2},m_{%
{\nu }_3})\cdot U_\nu ^T$}. Note that both $M_\ell $ and $M_\nu $ have only
three parameters each. As a result, both $U_\ell $ and $U_\nu $ will be
entirely fixed by charged lepton and neutrino mass ratios. Due to the
observed strong hierarchy in the charged lepton mass spectrum, the
parameters $(a,b)$ will be close to zero. From the form of $M_\ell $ in Eq.(%
\ref{eq12}) one can derive exact expressions for the phases $(a,b)$ in terms
of charged lepton mass ratios. In leading order, one obtains:

\begin{equation}
\label{eq14}
\begin{array}{c}
\begin{array}{ccc}
a=3\sqrt{3}\ {\frac{\sqrt{m_em_\mu }}{m_\tau }}~ & \ ;\ \ \  & b={\frac 92\ }%
{\frac{m_\mu }{m_\tau }} 
\end{array}
~ 
\end{array}
\end{equation}

In the neutrino sector, it was already pointed out that the square mass
differences necessary to explain the solar neutrino data, together with the
requirement that neutrinos constitute the hot dark matter, lead to highly
degenerate neutrinos. The matrix $M_\nu $ in Eq.(\ref{eq12}) leads to
threefold degeneracy for $\alpha =\beta =2\pi /3$. Thus, it is useful to
introduce the two small parameters $\epsilon _{_{32}},\epsilon _{_{21}}$,
defined by:

\begin{equation}
\label{eq15}
\begin{array}{ccc}
\alpha =\frac{2\pi }3-\epsilon _{_{32}}-\epsilon _{_{21}} & \ ;\ \ \  & 
~\beta ={\frac{2\pi }3}+2\epsilon _{_{32}}
\end{array}
\end{equation}
It is clear from this parametrization, that in the limit $\epsilon _{_{21}}=0
$, one has for the phase $\beta =-2\alpha $ (mod $2\pi $), and therefore one
obtains in this limit the mass matrix $M_\nu ^{{\rm I\!I\!}}$ of Eq.(\ref
{eq11}), with two exactly degenerate masses. Using Eq.(\ref{eq12}), one
calculates the eigenvalues $\lambda _i$ of $H_\nu =M_\nu M_\nu ^{\dagger }\
/\ 3{c_\nu }^2$ as functions of $\alpha $ and $\beta $, 
\begin{equation}
\label{eq15a}
\begin{array}{lll}
\lambda _1=1-y;\quad  & \lambda _2=\frac{2+y-\sqrt{8x^2+y^2}}2;\quad  & 
\lambda _3=\frac{2+y+\sqrt{8x^2+y^2}}2
\end{array}
\end{equation}
where $y=(2\cos (\alpha )+1)/3$ and $x=(1/3)\ [3+2\cos (\alpha )+2\cos
(\beta )+2\cos (\alpha +\beta )]^{1/2}$. Then from Eq.(\ref{eq15a}) one can
express $\epsilon _{_{32}},\epsilon _{_{21}}$ in terms of mass ratios. In
leading order, one obtains:

\begin{equation}
\label{eq16}
\begin{array}{ccc}
\epsilon _{_{32}}={\frac 1{\sqrt{3}}}{\frac{\Delta m_{32}^2}{m_3^2}} & \ ;\
\ \  & \epsilon _{_{21}}={\frac{3\sqrt{3}}4}{\frac{\Delta m_{21}^2}{m_3^2}} 
\end{array}
\end{equation}

In order to evaluate the leptonic mixing matrix $V$, it is convenient to
start by making a weak-basis (WB) transformation defined by:

\begin{equation}
\label{eq17}
\begin{array}{ccc}
H_\ell \equiv \frac 1{\ 3{c}_\ell ^2}\ M_\ell M_\ell ^{\dag }\  & 
\longrightarrow & H_\ell ^{\prime }=F^{\dag }\cdot H_\ell \cdot F \\  
&  &  \\ 
H_\nu \equiv \frac 1{\ 3{c_\nu }^2}\ M_\nu M_\nu ^{\dag } & \longrightarrow
& H_\nu ^{\prime }=F^{\dag }\cdot H_\nu \cdot F 
\end{array}
\end{equation}
where

\begin{equation}
\label{eq18}F=\left[ 
\begin{array}{lll}
\frac 1{\sqrt{2}} & \frac{-1}{\sqrt{6}} & \frac 1{
\sqrt{3}} \\ \frac{-1}{\sqrt{2}} & \frac{-1}{\sqrt{6}} & \frac 1{
\sqrt{3}} \\ 0 & \frac 2{\sqrt{6}} & \frac 1{\sqrt{3}} 
\end{array}
\right] 
\end{equation}

Obviously, after this WB transformation the charged leptonic weak current
remains diagonal, real, i.e., $V={1\>\!\!\!{\rm I}}$. This WB transformation
corresponds to writing $H_\ell $ in a ``heavy basis''. Due to the strong
hierarchy of the charged lepton masses, in this ``heavy basis'' all other
elements of $H_\ell ^{\prime }$ are small, compared to the element $(3,3)$.
The Hermitian matrix $H_\ell ^{\prime }$ can then be diagonalized, and one
obtains in leading order:

\begin{equation}
\label{eq19}
\begin{array}{lll}
|U_{12}^\ell |=\sqrt{{\frac{m_e}{m_\mu }}} & \ ;\ \ \  & |U_{13}^\ell |= 
{\frac{\sqrt{2m_em_\mu }}{m_\tau }} \\  &  &  \\ 
|U_{31}^\ell |={\frac 3{\sqrt{2}}\ \frac{\sqrt{m_em_\mu }}{m_\tau }} & \ ;\
\ \  & |U_{23}^\ell |={\frac 1{\sqrt{2}}\ }{\frac{m_\mu }{m_\tau }} 
\end{array}
\end{equation}

In the neutrino sector, one has a drastically different situation. After the
WB transformation of Eq.(\ref{eq17}), the first neutrino exactly decouples
from the other two, i.e., $(H_\nu ^{\prime })_{12}=(H_\nu ^{\prime })_{13}=0$%
.

Furthermore, the mixing between the second and third neutrino state is large
and, to leading order, independent of neutrino mass ratios, one obtains:

\begin{equation}
\label{eq20}U_\nu =\left[ 
\begin{array}{ccc}
1 & 0 & 0 \\ 
0 & {\frac{\omega ^{\star }-\omega }3} & {\frac{\sqrt{2}(\omega ^{\star }-1)}%
3} \\ 0 & {\frac{\sqrt{2}(\omega ^{\star }-1)}3} & {\frac{1-\omega }3} 
\end{array}
\right] 
\end{equation}
where $\omega =e^{i2\pi /3}$. In Eq.(\ref{eq20}) the zeros are exact yet we
have omitted in the other entries terms of the order $\Delta m_{21}^2/\Delta
m_{32}^2$. Therefore, the leptonic mixing matrix is given, in leading order,
by:

\begin{equation}
\label{eq21}|V|=\left[ 
\begin{array}{ccc}
1 & \sqrt{{\frac{m_e}{m_\mu }}} & \frac{\sqrt{2m_em_\mu }}{m_\tau } \\ \sqrt{%
{\frac{m_e}{3m_\mu }}} & \frac 1{\sqrt{3}} & \frac{\sqrt{2}}{\sqrt{3}} \\ 
\sqrt{\frac{2m_e}{3m_\mu }} & \frac{\sqrt{2}}{\sqrt{3}} & \frac 1{\sqrt{3}} 
\end{array}
\right] 
\end{equation}

\section{\bf Confronting the data}

Neutrino oscillations $\cite{re9}$ have to be taken into account since
neutrino experiments are discussed in terms of neutrino weak eigenstates
rather than physical states. With the notation $\nu _{\alpha (\beta )}$ for
weak eigenstates and $\nu _{i(j)}$ for mass eigenstates, the probability of
finding $\nu _\beta $ at time $t$ having started with $\nu _\alpha $ at $t=0$
neglecting the effect of CP violation in the leptonic sector (real $V$) is
given by:

\begin{equation}
\label{eq22}
\begin{array}{c}
P(\nu _\alpha \rightarrow \nu _\beta )\equiv <\nu _\beta |\nu _\alpha
(t)>\cdot <\nu _\beta |\nu _\alpha (t)>^{\star }= \\  
\\ 
=\delta _{\alpha \beta }-4\ {\sum\limits_{i<j}\ }V_{\alpha i}V_{\beta
i}V_{\alpha j}V_{\beta j}\cdot \sin ^2\left[ {\frac{\Delta m_{ji}^2}4}{\frac
LE}\right] 
\end{array}
\end{equation}
where $E$ is the neutrino energy, $L$ is the distance travelled by the
neutrino between the source and the detector, $V$ is the leptonic mixing
matrix given by Eq.(\ref{eq13}), and $\Delta m_{ji}^2$ is defined by:

\begin{equation}
\label{eq23}
\begin{array}{c}
\Delta m_{ji}^2=|m_j^2-m_i^2|
\end{array}
\end{equation}

The interpretation of the experimental data is presented in terms of two
flavour mixing, in this case Eq.(\ref{eq22}) reduces to

\begin{equation}
\label{eq24}
\begin{array}{c}
P(\nu _\alpha \rightarrow \nu _\beta )=\delta _{\alpha \beta }-\sin
^22\theta \cdot \sin ^2\left[ {\frac{\Delta m_{ji}^2}4}{\frac LE}\right] 
\end{array}
\end{equation}
hence the meaning of the experimental bounds presented in section 1.

The translation of the bounds into the three flavour approach is quite
simple for the experimental limits imposed on $\Delta m_{ji}^2$, with $%
\Delta m_{21}^2$ in the solar range and $\Delta m_{32}^2$ in the atmospheric
range. In the case of the atmospheric anomaly it is clear that we can
disregard the term in $\sin ^2[(\Delta m_{21}^2/4)(L/E)]$ and we can
approximately write

\begin{equation}
\label{eq25}
\begin{array}{c}
\sin ^2\left[ {\frac{\Delta m_{31}^2}4}{\frac LE}\right] \simeq \sin
^2\left[ {\frac{\Delta m_{32}^2}4}{\frac LE}\right] \equiv S 
\end{array}
\end{equation}
so that

\begin{equation}
\label{eq26}
\begin{array}{c}
1-P(\nu _\mu \rightarrow \nu _\mu )=P(\nu _\mu \rightarrow \nu _e)+P(\nu
_\mu \rightarrow \nu _\tau )=4\
(V_{23}V_{23}V_{21}V_{21}+V_{22}V_{22}V_{23}V_{23})\ S 
\end{array}
\end{equation}
and we identify

\begin{equation}
\label{eq27}
\begin{array}{c}
\sin ^22\theta _{{\rm {atm}}%
}=4(V_{23}V_{23}V_{21}V_{21}+V_{22}V_{22}V_{23}V_{23}) 
\end{array}
\end{equation}

In the case of the solar neutrino anomaly the range $L/E$ is such that $S$
in Eq.(\ref{eq25}) can be averaged to ${\frac 12}$ and we obtain

\begin{equation}
\label{eq28}
\begin{array}{c}
1-P(\nu _e\rightarrow \nu
_e)=4(V_{11}V_{11}U_{13}U_{13}+V_{12}V_{12}V_{13}V_{13})\cdot {\frac 12}%
+4V_{11}V_{11}V_{12}V_{12}\sin ^2[{\frac{\Delta m_{32}^2}4}{\frac LE}] 
\end{array}
\end{equation}
in our numerical example the first term of this equation is small so, as a
result, we can identify

\begin{equation}
\label{eq29}
\begin{array}{c}
\sin ^22\theta _{{\rm {sol}}}=4V_{11}V_{11}V_{12}V_{12} 
\end{array}
\end{equation}

Our ansatz can perfectly fit the experimental bounds as is shown by the
following example.

We choose as input the masses for the charged leptons, 
\begin{equation}
\label{eq29a}
\begin{array}{ccc}
m_e=0.511\ MeV,\quad & m_\mu =105.7\ MeV,\quad & m_\tau =1777\ MeV 
\end{array}
\end{equation}
which correspond to the phases $a=0.0214$ and $b=0.2662$ of Eq.(\ref{eq12}).
For the neutrino sector, we choose, 
\begin{equation}
\label{eq29b}
\begin{array}{ccc}
m_{\nu _3}=2\ eV,\quad & \Delta m_{21}^2=9.2\times 10^{-6\ }eV^2,\quad & 
\Delta m_{32}^2=5.0\times 10^{-3}\ eV^2 
\end{array}
\end{equation}
This fixes the values of the parameters $\epsilon _{_{21}}=3\times 10^{-6}$
and $\epsilon _{_{32}}=7.2\times 10^{-4}$. With the above input we obtain
for the leptonic mixing matrix $V$, without approximations:

\begin{equation}
\label{eq30}|V|=\left[ 
\begin{array}{lll}
0.9976 & 0.0692 & 0.0058 \\ 
0.0463 & 0.6068 & 0.7935 \\ 
0.0518 & 0.7918 & 0.6085 
\end{array}
\right] 
\end{equation}
Making use of Eqs.(\ref{eq27}) and (\ref{eq29}), Eq.(\ref{eq30}) translates
into

\begin{equation}
\label{eq31}
\begin{array}{c}
\sin ^22\theta _{{\rm {atm}}}=0.933 
\end{array}
\end{equation}
and

\begin{equation}
\label{eq31a}
\begin{array}{c}
\sin ^22\theta _{{\rm {sol}}}=0.019 
\end{array}
\end{equation}
These results are in agreement with the present experimental data. In this
example the three physical neutrinos under consideration all have masses
close to $2\ eV$. Note that the leptonic mixing matrix $V$, which differs
significantly from the observed quark mixing matrix, was obtained having as
input the charged lepton and neutrino masses, with no further parameters.

%%%%%%%%%%%%%%%%%%% summary %%%%%%%%%%%%%%%%%%%%%%%%%%%%%%%

\section{\bf Concluding remarks}

We have suggested an ansatz for the neutrino and charged lepton mass
matrices, within the framework of universality of strength of Yukawa
couplings. The ansatz has the same structure of phases for the neutrino and
charged lepton mass matrices, with the only non-vanishing phases along the
diagonal. Both $M_\nu $ and $M_\ell $ have only three parameters each, two
phases and an overall real constant. These parameters are completely fixed
by the value of charged lepton and neutrino masses, which implies that the
ansatz is highly predictive, with full calculability of the leptonic mixing
matrix, i.e., $V$ is completely fixed by charged lepton and neutrino mass
ratios. We have shown that the ansatz naturally leads on the one hand to
small mixing among ${\nu }_e$ and ${\nu }_\mu $, ${\nu }_\tau $ associated
to $\Delta m_{21}^2\sim 10^{-5}\ eV^2$ thus explaining the solar neutrino
deficit in the non-adiabatic MSW solution, and on the other hand leads to
large mixing between $\nu _\mu $ and $\nu _\tau $, as required to explain
the atmospheric neutrino anomaly.

We find our results specially appealing, since one has an unified view of
all Yukawa couplings, i.e., both quark and lepton Yukawa couplings have
universal strength, with the flavour dependence being all contained in their
phases. Furthermore, both in the lepton and quark sectors $\cite{re8}$ one
has simple ans\"atze within USY, whose distinctive feature is having a
number of independent parameters equal to the number of elementary fermion
masses. As a result, one has highly predictive schemes, with the fermion
mixings expressed in terms of fermion mass ratios with no free parameters.

%%%%%%%%%%%%%%%%%% acknowlegement %%%%%%%%%%%%%%%%%%%%

\section*{\bf Acknowledgments}

We would like to thank M. Tanimoto for enlightening discussions during the
preparation of this work.

%%%%%%%%%%%%%%%%%%%% bibliography %%%%%%%%%%%%%%%%%%%%

%%%%%%%%%%%%%%%%%%%%%%%%%%%%%%%%%%%%%%%%%%%%%%%%%%%%

\end{document}